\documentclass[twocolumn,english,aps,prb,showpacs,superscriptaddress]{revtex4}
\usepackage[T1]{fontenc}
\usepackage[latin9]{inputenc}
\usepackage{verbatim}
\usepackage{graphicx}

\usepackage{babel}

\begin{document}

\title{A First-Principles Study of Zinc Oxide Honeycomb Structures}

\author{M. Topsakal}
\affiliation{UNAM-Institute of Materials Science and
Nanotechnology, Bilkent University, Ankara 06800, Turkey}
\author{S. Cahangirov}
\affiliation{UNAM-Institute of Materials Science and
Nanotechnology, Bilkent University, Ankara 06800, Turkey}
\author{E. Bekaroglu}
\affiliation{UNAM-Institute of Materials Science and
Nanotechnology, Bilkent University, Ankara 06800, Turkey}
\author{S. Ciraci}
\affiliation{UNAM-Institute of Materials Science and
Nanotechnology, Bilkent University, Ankara 06800, Turkey}
\email{ciraci@fen.bilkent.edu.tr} \affiliation{Department of
Physics, Bilkent University Ankara 06800, Turkey}
\date{\today}

\begin{abstract}
We present a  first-principles study of the atomic, electronic and
magnetic properties of two dimensional (2D), single and bilayer
ZnO in honeycomb structure and its armchair and zigzag
nanoribbons. In order to reveal the dimensionality effects, our
study includes also bulk ZnO in wurtzite, zincblende and hexagonal
structures. The stability of 2D ZnO, its nanoribbons and flakes are
analyzed by phonon frequency, as well as by finite temperature
ab-initio molecular dynamics calculations. 2D ZnO in honeycomb
structure and its armchair nanoribbons are nonmagnetic
semiconductors, but acquire net magnetic moment upon the creation
of zinc vacancy defect. Zigzag ZnO nanoribbons are ferromagnetic
metals with spins localized at the oxygen atoms at the edges and
have high spin polarization at the Fermi level. However, they
change to nonmagnetic metal upon termination of their edges with
hydrogen atoms. From the phonon calculations, the fourth acoustical mode specified as twisting
mode is also revealed for armchair nanoribbon. Under tensile
stress the nanoribbons are deformed elastically maintaining
honeycomb like structure, but yield at high strains. Beyond
yielding point honeycomb like structure undergo a structural
change and deform plastically by forming large polygons. The
variation of the electronic and magnetic properties of these
nanoribbons have been examined under strain. It appears that
plastically deformed nanoribbons may offer a new class of
materials with diverse properties.
\end{abstract}

\maketitle

\section{introduction}

Graphene, a monolayer layer of carbon atoms in honeycomb
structure, is offering exceptional
properties\cite{novo,zhang,berger} which may lead to important
applications in various fields. Normally, two dimensional (2D)
graphene is semimetallic and its electrons and holes behave like a
massless Dirac fermion. Whereas 2D boron-nitride
(BN),\cite{bn-synthesis_bn-insulator}  Group III-V analogue of
graphene, in ionic honeycomb structure is a wide band gap
semiconductor. Unusual properties of graphene and BN nanoribbons
have been revealed extensively in recent
papers.\cite{gribbon1,gribbon2,gribbon3,gribbon4,bnribbon1,bnribbon2,bnribbon3}
More recently, based on state-of-the art first principles
calculations it was predicted that Si and Ge\cite{seymur}, even
binary compounds of Group IV elements and III-V
compounds\cite{hasan} can form 2D stable monolayer honeycomb
structures. Earlier studies on ZnO and its nanowires gave first
indications that graphitic ZnO can
exists.\cite{claeyssens,kulkarni} Very thin
nanosheets,\cite{nanosheet} nanobelts,\cite{nanobelt}
nanotubes\cite{nanotube} and nanowires\cite{nanowire} of ZnO have
already been synthesized. Two monolayer thick ZnO(0001) films
grown on Ag(111) were reported.\cite{bilayer}

Because of its wide band gap of $\sim$3.3 eV and large exciton
binding energy of 60 meV leading to vast optoelectronic
applications including light-emitting diodes and solar
cells,\cite{diode,solar-cell} ZnO has been the subject of various
researches. It was reported that the vacancy defects can be
intentionally created by electron irradiation
method.\cite{annihilation} It has been also reported that Zn
vacancy induces ferromagnetism in ZnO thin films and nanowires
without any need of doping with transition-metal atoms.\cite{jena}
These magnetic properties might provide superior advantages in
biomedical applications because of non-toxic nature of ZnO as
opposed to transition metal ions.

In this paper a comprehensive study of the atomic, electronic and
magnetic properties of monolayer, bilayer and nanoribbons of II-VI
ionic ZnO compound in honeycomb structures are carried out using
first-principles calculations. In order to reveal the
dimensionality effects, we started with the energetics and
electronic energy bands of ZnO in different 3D bulk crystalline
structures and compared them with those of 2D ZnO honeycomb
structure. Our analysis based on phonon dispersions and finite
temperature ab-initio molecular dynamics calculations provides
evidence for the stability of free-standing 2D monolayer, bilayer
and quasi 1D nanoribbons of ZnO honeycomb structures. These
structures can be in different local minima on the
Born-Oppenheimer surface, in spite of the fact that they are not
synthesized yet. We found that 2D monolayer and bilayer ZnO are
nonmagnetic semiconductors, but attain magnetic properties upon
creation of Zn-vacancy defect. ZnO nanoribbons exhibit interesting
electronic and magnetic properties depending on their orientation.
While armchair ZnO nanoribbons are nonmagnetic semiconductors with
band gaps varying with their widths, bare zigzag nanoribbons are
ferromagnetic metals. These electronic and magnetic properties
show dramatic changes under elastic and plastic deformation.
Hence, ZnO nanoribbons can be functionalized by plastic
deformation. Results obtained in this study indicates that 2D
monolayer and bilayer ZnO honeycomb structures and quasi 1D
armchair and zigzag nanoribbons display unusual electronic,
magnetic and mechanical properties, which hold the promise of
interesting technological applications.

\section{Model and Methodology}

We have performed first-principles plane wave calculations within
Density Functional Theory (DFT) using PAW potentials.\cite{paw}
The exchange correlation potential has been approximated by
Generalized Gradient Approximation (GGA) using PW91 \cite{pw91}
functional both for spin-polarized and spin-unpolarized cases.
Recently, spin-polarized calculations within DFT have been used
successfully to investigate magnetic properties of vacancy defects in
2D honeycomb structures. Also interesting spintronic properties of
nanoribbons have been revealed using spin-polarized DFT.\cite{gribbon2}
The success of spin-polarized DFT calculations has been discussed in
Ref{\onlinecite{zeller}.

All structures have been treated within supercell geometry using
the periodic boundary conditions. A plane-wave basis set with
kinetic energy cutoff of 500 eV has been used. The interaction
between ZnO monolayers in adjacent supercells is examined as a
function of their spacing. Since the total energy per cell has
changed less than 1 meV upon increasing the spacing from 10 \AA~
to 15 \AA~, we used the spacing of $\sim10$ \AA~ in the
calculations. In the self-consistent potential and total energy
calculations the Brillouin zone (BZ) is sampled by, respectively
(15x15x15), (25x25x1) and (25x1x1) special \textbf{k}-points for
3D bulk, 2D honeycomb and 1D (nanoribbons) ZnO. This sampling is
scaled according to the size of superlattices. For example, BZ was
sampled by (3$\times$3$\times$1) special \textbf{k}-points for
defect calculations using (7$\times$7) supercell of 2D ZnO
honeycomb structure. All atomic positions and lattice constants
are optimized by using the conjugate gradient method, where the
total energy and atomic forces are minimized. The convergence for
energy is chosen as 10$^{-5}$ eV between two steps, and the
maximum Hellmann-Feynman forces acting on each atom is less than
0.02 eV/\AA{} upon ionic relaxation. The pseudopotentials having
12 and 6 valence electrons for the Zn (Zn: $4s^{2}$ $3d^{10}$) and
O ions (O: $2s^{2}$ $2p^{4}$) are used. Numerical plane wave
calculations are performed by using VASP
package.\cite{vasp1,vasp2} While all numerical calculations of
structure optimization, electronic energy and phonon dispersions
are carried out within GGA using VASP,\cite{vasp1,vasp2} some of
the calculations are checked also by using PWSCF\cite{pwscf} and
SIESTA\cite{siesta} softwares. Therefore all pertaining
discussions are based on the results obtained by using VASP
software unless it is stated otherwise.

Since DFT within GGA underestimates the band gap,
frequency-dependent \textit{GW$_{0}$} calculations are carried
out.\cite{gw} Screened Coulomb potential, W, is kept fixed to
initial DFT value W$_{0}$ and Green's function, G, is iterated
five times. Various tests regarding vacuum separation, kinetic energy
cut-off potential, number of bands, \textbf{k}-points and grid
points are made. Final results of \textit{GW$_{0}$} corrections are
obtained using (12$\times$12$\times$1) \textbf{k}-points in BZ,
$20$~\AA~ vacuum separation, 400 eV cut-off potential, 160 bands and 64
grid points. In addition to \textit{GW$_{0}$}, we performed also \textit{GW}
calculations in order to make comparison with earlier available
studies. While \textit{GW$_{0}$} corrections are successfully applied to 3D
and 2D ZnO, its application to quasi 1D nanoribbons is hindered by
large number of atoms.

\section{3D bulk and 2D honeycomb ZnO crystal}

We first consider 3D bulk ZnO, which are in wurtzite (wz-ZnO),
zincblende (zb-ZnO) and hexagonal (h-ZnO, or graphite-like)
crystals. Atoms in wz- and zb-ZnO structures are four fold
coordinated through tetrahedrally directed
\textit{$sp^3$-}orbitals, whereas the atoms in h-ZnO crystal are
three fold coordinated through \textit{$sp^2$-}orbitals. Wurtzite
structure is found to be the thermodynamically most stable phase
of ZnO. The cohesive energy per Zn-O pair is calculated by using
the expression,

\begin{equation}\label{equ:binding}
E_C={E_{T}[ZnO]} - E_{T}[Zn] - E_{T}[O]
\end{equation}

in terms of the total energy of the optimized crystal structure of
ZnO, $E_{T}[ZnO]$ per Zn-O pair, the total energies of free Zn and
O atoms $E_{T}[Zn]$ and $E_{T}[O]$. The equilibrium cohesive
energies of wz-ZnO, zb-ZnO and h-ZnO structures are found to be
8.934, 8.919 and 8.802 eV per Zn-O pair, respectively. For
wz-ZnO crystal, the hexagonal lattice constants of the optimized
structure in equilibrium are $a = 3.280$ \AA{}, $c/a = 1.616$. The
deviation of $c/a$ from the ideal value of 1.633 imposes a slight
anisotropy in the lengths of tetrahedrally directed Zn-O bonds.
While the length of three short bonds is 2.001 \AA, the fourth
bond is slightly longer and has the length of 2.007 \AA . The
zincblende structure in $T_d$ symmetry has cubic lattice constant
$a = 3.266$ \AA{} and four tetrahedrally coordinated Zn-O bonds
having uniform length, $d=2.001$ \AA{} . The h-ZnO structure has
hexagonal lattice constants $a = 3.448$ \AA{}, $c/a = 1.336$ and
$d=1.990$. The structural parameters of these three bulk ZnO
crystals are shown in Fig.~\ref{fig:Figure-Bulks}. The lattice
constant of wz-ZnO, $a$ ($c/a$) is measured between 3.247 (1.6035)
\AA~ and 3.250 (1.602) \AA~ using different
methods.\cite{bulk_zno} The calculations based on ab-initio LCAO
method with all-electron Gaussian type basis set predict $a$=3.286
\AA~ and $c/a$=1.595.\cite{allelect} All our results related with
the structural parameters are in good agreement with the
experimental and theoretical values within the average error of
$\sim$ 1 \%.\cite{bulk_zno,allelect}

\begin{figure}
\begin{center}
\includegraphics[width=7.2cm]{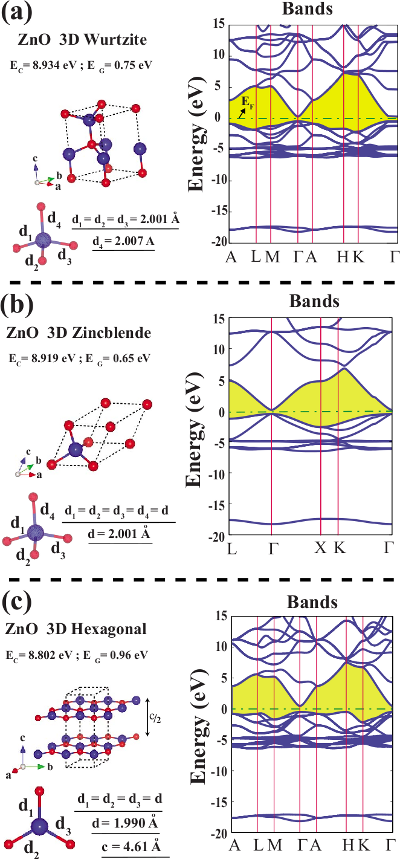}
\caption{(Color online) Cohesive energies $E_{C}$ per Zn-O pair,
band gaps $E_{G}$, Zn-O bonds, atomic and energy band structures
of 3D wurtzite (a), zincblende (b), and hexagonal (c) crystals of
bulk ZnO. Red-small and blue-large balls correspond to O and Zn
atoms, respectively. The gap between valence and conduction bands
are shaded and the zero of energy is set at the Fermi level
$E_{F}$. All structures including lattice constants are fully
optimized.} \label{fig:Figure-Bulks}
\end{center}
\end{figure}

The optimized atomic structure and corresponding electronic band
structure of 3D ZnO crystals are presented in
Fig.~\ref{fig:Figure-Bulks}. All wz-ZnO, zb-ZnO and h-ZnO crystals
are direct band gap semiconductors with calculated band gaps being
$E_{G}$=0.75, 0.65 and 0.96 eV, respectively. Highest valence band
has O-\textit{2p} character; the states of the lowest conduction
band is formed from Zn-\textit{4p} and Zn-\textit{4s} orbitals.
Valence band consists of two parts separated by a wide intra band
gap. The lower part at $\sim$ -18 eV is projected mainly to O-$2s$
orbitals. The upper part is due to mainly Zn-$3d$ and O-$2s$
orbitals.  The differences in the band structure of different
three 3D crystals become pronounced in the lower part of the
conduction band. It should be noted that the band gaps of bulk ZnO
is highly underestimated by DFT calculations.\cite{hafner} The
experimentally measured band gap of wz-ZnO is $\sim$3.37 eV at
room temperature.\cite{bulk_zno} We performed \textit{GW$_{0}$}
calculations to correct the band gaps calculated within GGA. Our
results for  wz-ZnO, zb-ZnO and h-ZnO are, respectively, 3.29,
3.04 and 3.32 eV. As for earlier studies, calculations with \textit{GW}
corrections reported a band gap of 3.59 eV for zb-ZnO.\cite{zb_gw}
All electron LAPW calculations predicted the band gap of wz-ZnO
0.77 eV using LDA, which is corrected by \textit{GW} calculations to 2.44
eV.\cite{hamada}

Charge transfer from Zn atoms to O atoms is a measure of the
ionicity of ZnO crystal. We calculated the amount of charge on
constituent Zn and O atoms in 3D crystals by performing the
L\"{o}wdin analysis\cite{lowdin} in terms of the projection of
plane-waves into atomic orbitals. By subtracting the valencies of
free Zn and O atoms from the calculated charge values on the same
atoms in 3D crystals we obtain the charge transfer $\delta q$ from
Zn to O. The calculated value of charge transfer for both wz-ZnO,
zb-ZnO and h-ZnO is found to be 1.41 electrons. The values of
charge transfer calculated with the Bader analysis\cite{bader} are
$\delta q =1.22$, $\delta q=1.17$ and $\delta q=1.20$ electrons
for wz-ZnO, zb-ZnO and h-ZnO, respectively. Same analysis
performed with local basis set using SIESTA\cite{siesta} yields
significantly lower values of charge transfer, $\delta q =0.90$,
$\delta q=0.88$ and $\delta q=0.91$ electrons for wz-ZnO, zb-ZnO
and h-ZnO, respectively. This analysis clearly indicates that a
significant amount of charge is transferred from low
electronegative Zn atom to high electronegative O atom. However
the values of $\delta$q may scatter owing to the ambiguities in
placing boundary between Zn and O in crystalline structure.

\begin{figure}
\begin{center}
\includegraphics[width=8cm]{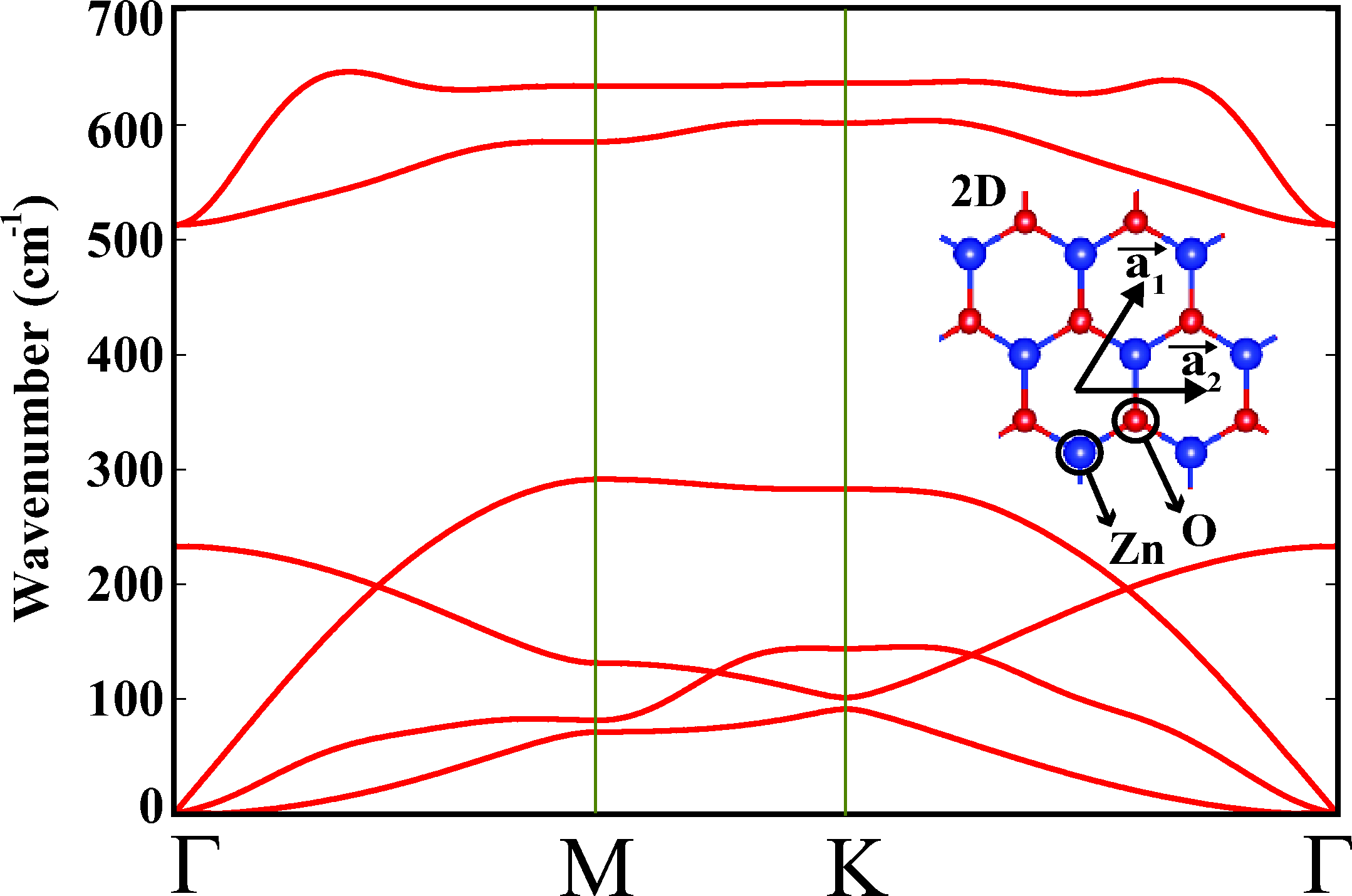}
\caption{(Color online) Phonon dispersion curves calculated by
force constant method for 2D monolayer ZnO. Atomic structure and
hexagonal lattice vectors are shown by inset.}
\label{fig:Figure-2d-Phonon}
\end{center}
\end{figure}

\section{2D ZnO Honeycomb Structure}

The structure of monolayer ZnO is optimized using periodically
repeating supercell having hexagonal lattice in 2D and the spacing
of 10 \AA~ between ZnO planes. The optimized structure was planar
and the magnitudes of the Bravais vectors of the hexagonal lattice
are found to be $a_{1}= a_{2} = 3.283$ \AA{}, and the Zn-O bond,
$d=$1.895 \AA{} (see Fig.~\ref{fig:Figure-2d-Phonon}). In contrast
to 2D puckered honeycomb structures of Si, Ge and compounds of III
and V group elements both lying below the first row, any honeycomb
structure including one element from the first row form planar
honeycomb structure like graphene, BN and SiC.\cite{hasan} 2D
monolayer of ZnO forming planar honeycomb structure is not an
exception. The calculated structural parameters are significantly
larger than those of graphene and BN honeycomb structure due to
fact that Zn has larger radius\cite{radius} than that of B,C,N and
O atoms. The planar structure of 2D ZnO is tested by displacing Zn
and O atoms arbitrarily from their equilibrium positions by 0.5
\AA~ and subsequently by re-optimizing the structure. Upon
optimization the displaced atoms have recovered their original
positions in the same plane. It should be noted that, the length
of Zn-O bonds of 2D ZnO honeycomb structure is smaller than that
in the 3D bulk (wz, zb) crystals, since $sp^2$ bonding in the
former is stronger than the tetrahedrally coordinated $sp^3$
bonding in the latter. Similar trend is also found in C, BN and
SiC honeycomb structures. The interaction between ZnO planes
appears to slightly weaken the Zn-O bonds of h-ZnO crystal. As a
result the length of the Zn-O bonds becomes larger than that of 2D
ZnO honeycomb structure. The cohesive energy of 2D monolayer of
ZnO is calculated to be 8.419 eV per Zn-O pair. This energy is 0.5
eV smaller than that of 3D wz-ZnO.

\begin{figure}
\begin{centering}
\includegraphics[width=7.8cm]{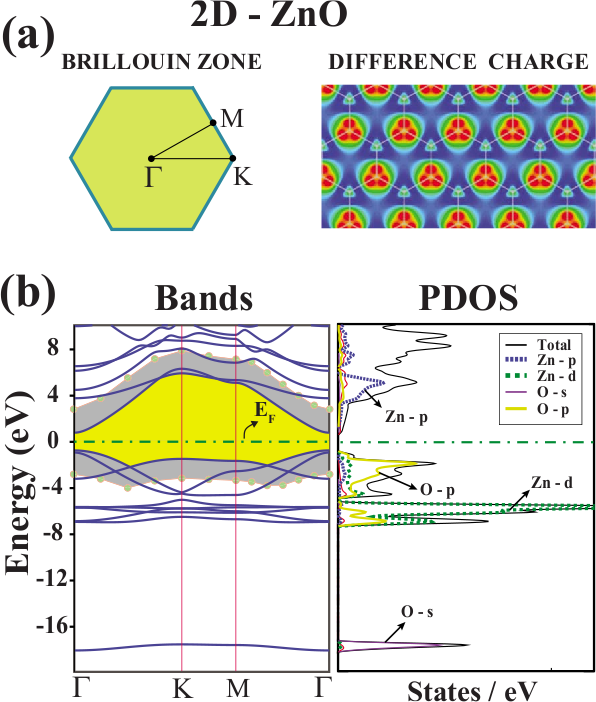}
\par\end{centering}
\caption{(Color online) Electronic structure of the 2D monolayer
of ZnO. (a) Brillouin zone  corresponding to 2D hexagonal lattice
and difference charge density, $\Delta \rho$. (b) Energy band
structure and density of states projected to the selected orbitals
(PDOS) of constituent atoms. The band gap is shaded (painted in
yellow) and the zero of energy is set at the Fermi level. Band
edges corrected by \textit{GW$_{0}$} are indicated by small ( light blue)
circles. The band gap enlarged after \textit{GW$_{0}$} correction is shaded
(painted light blue).} \centering{}\label{fig:Figure-2D}
\end{figure}

\subsection{Atomic structure and stability analysis}

It should be noted that 2D honeycomb structure determined by the
structure optimization using conjugate gradient method may not be
necessarily stable. One needs more stringent tests to assure the
stability of this truly 2D structure. As a matter of fact, it was
argued earlier that any crystal structures consisting of a truly
2D monolayer crystals cannot be
stable.\cite{stability1,stability2} Here, we summarize our
analysis on the stability of planar 2D hexagonal structure of ZnO
using calculation of phonon dispersion curves as well as ab-initio
finite temperature molecular dynamics calculations.

We calculated optical and acoustical branches of phonon frequency
using the density functional perturbation theory (DFPT) with plane
wave method as implemented in PWSCF software and the force
constant method\cite{alfe} with forces calculated using VASP. The
former method includes the polarization effects so that the
splitting of longitudinal and transverse optical modes at the
$\Gamma$-point (namely LO-TO splitting) is obtained. The force
constant method cannot yield the LO-TO splitting and is also very
sensitive to the mesh size in calculating forces under atomic
displacements and hence in setting up the dynamical matrix. In
fact, the imaginary frequencies of ZA branch (so called out of
plane acoustical branch) occur around the $\Gamma$-point as an
artifact of insufficient mesh size. However, all imaginary
frequencies around the $\Gamma$-point (corresponding to an
instability to be induced by acoustical waves with large
$\lambda$) are removed by using very fine mesh. Earlier, imaginary
frequencies of ZA modes near the $\Gamma$-point calculated for 2D
honeycomb structures of Ge and III-V compounds inducing similar
fortuitous instabilities for large $\lambda$ were also removed by
using finer mesh.\cite{hasan}

In Fig.~\ref{fig:Figure-2d-Phonon} we present phonon dispersion
calculated for 2D monolayer of ZnO honeycomb structure using force
constant method, where atomic forces are calculated by VASP.
Calculated phonon modes all being positive across the BZ strongly
support that 2D monolayer of ZnO is a stable structure
corresponding to a local minimum on the Born-Oppenheimer surface.
Our phonon dispersion curves are in agreement with those
calculated in Ref[\onlinecite{phonon}].

Furthermore, we have tested the stability of 2D ZnO monolayer
using finite temperature ab initio molecular dynamics (MD)
calculations with time steps of 2x10$^{-15}$ seconds. To lift the
constraints to be imposed by a small unit cell, we considered a
(7$\times$7) supercell of the 2D hexagonal ZnO and raised its
temperature from 0 K to 750 K in 0.1 ps. Then, we have kept the
temperature of the system around 750 K for 2.5 ps. While all these
calculations resulted in minor deformations, the honeycomb
structure was not destroyed. It should be noted that these
calculations may not be conclusive, since 2.5 ps cannot be
sufficient to represent all the statistics, but this picture is
the best one can see from the exiguous window limited by the
computational cost imposed by ab-initio MD method. In the
following sections we will present additional arguments related
with electronic structure, which further corroborate the stability
of 2D ZnO structures.

\subsection{Electronic Structure}

The difference charge density and the electronic energy bands
together with the partial density of states are presented in
Fig.~\ref{fig:Figure-2D}. Contour plots of total charge indicate
high density around O atoms. The difference charge density is
calculated by subtracting charge densities of free Zn and O atoms
from the total charge density of 2D ZnO, i.e. $\Delta
\rho=\rho_{ZnO}-\rho_{Zn}-\rho_{O}$. High density contour plots
around O atoms protruding towards the Zn-O bonds indicate charge
transfer from Zn to O atoms. This way the Zn-O bond acquires an
ionic character. The charge transfer from Zn to O, $\delta$q is
analyzed by using different schemes. The charge transfer values
calculated by L\"{o}wdin method using PWSCF , Bader\cite{bader}
analysis using VASP, and local basis set analysis using SIESTA
are, respectively, $\delta q$=1.35, 1.18 and 0.87 electrons.
Interestingly, as compared to 3D bulk h-ZnO crystal, the charge
transfer from cation to anion of 2D monolayer ZnO is slightly
decreased. This is due to the change from $sp^{3}$ hybrid orbital
in h-ZnO to $sp^{2}$ hybrid orbital in honeycomb structure.

Two dimensional ZnO is a direct band gap semiconductor with a gap
value of 1.68 eV. However, the actual band gap is expected to be
larger. The bands are corrected using \textit{$GW_{0}$} method\cite{gw} and
the direct band gap at the $\Gamma$-point of BZ is found to be
5.64 eV. Much recently, it has been reported that the band gap of
2D-ZnO is calculated to be 3.57 eV with \textit{GW}
corrections.\cite{phonon} Similar to the bands of 3D bulk ZnO
crystals, the upper part of the upper valance band are derived
mainly from O\textit{-$2p$} orbitals, whereas the lower part has
Zn-$3d$ character. The bands at the edges of conduction and
valence bands along the $\Gamma$-$K$ direction are derived from
bonding and antibonding combination of O-$2p_z$ and Zn-$4p_z$
orbitals forming $\pi$- and $\pi^*$-states. The highest valence
band along the $\Gamma$-$K$ direction has mainly O-$2p_z$ but
small Zn-$4p_z$ orbital contribution, while the lowest conduction
band is composed mainly from Zn-$4p_z$ but small O-$2p_z$
orbitals. Small Zn-$4p_z$ contribution is also confirmed by PDOS.
Consequently, planar geometry of 2D monolayer of ZnO is expected
to be attained by the bonding combination of $p_z$-states. The
same situation occurs for 2D monolayer of BN honeycomb structure,
which is an ionic III-V compound with wide band gap between $\pi$-
and $\pi^*$-states. The planar stability graphene is also maintained by $\pi$-bonds.

\subsection{Vacancy Defects and Antisite}\label{vacancy}

It was shown that vacancies have  remarkable effects on electronic
and magnetic properties of 2D graphene and graphene
nanoribbons.\cite{esquinazi,Iijima,yazyev,guinea,brey2,topsakal_delik}
To the best of our knowledge, the effects of vacancies in
monolayer of ZnO have not been treated yet. It was reported that
the vacancy defects can be intentionally created by electron
irradiation method\cite{annihilation} on ZnO thin films. We
investigated the effects of Zn-, O-, Zn+O-divacancy and
O+Zn-antisite in a periodically repeating (5$\times$5) as well as
(7$\times$7) supercells. The vacancy-vacancy coupling in the
larger (7$\times$7) supercell was reduced significantly. Flat
bands in the band gap have charge density localized at the defect
site. The width of these flat bands can be taken as the measure of
the strength of the direct and indirect (via the hopping through
the orbitals in the lattice) vacancy-vacancy coupling. The largest
width of such a band is already small (less than 50 meV). The
average energy of these flat bands from the top of the valence
band can be taken as the localized states of an individual vacancy
defect. Our results obtained using a (7$\times$7) supercell are
presented in Fig.~\ref{fig:Figure-Vacancy} and discussed in the
rest of the section.

\begin{figure}
\begin{centering}
\includegraphics[width=7.8cm]{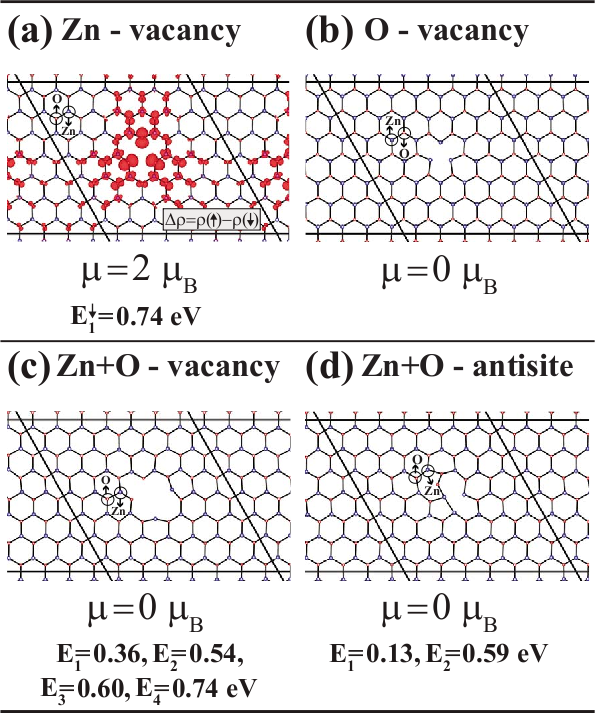}
\par\end{centering}
\caption{(Color online) Vacancy defects in a (7$\times$7)
supercell of the monolayer of ZnO. (a) Relaxed atomic structure
around the Zn-vacancy with isosurfaces of the difference charge
density of spin-up and spin-down states, $\Delta \rho =
\rho(\uparrow)=\rho(\downarrow)$. Energies $E_i$ of localized
states measured from the top of the valance band and the net
magnetic moment per supercell in Bohr Magneton $\mu_B$. Blue-large
and red-small balls indicate Zn and O atoms indicated by arrows.
(b) O-vacancy. (c) Zn+O divacancy. (d) O+Zn antisite where
neighboring O and Zn atoms are exchanged. The (7$\times$7)
supercells each including one type of various vacancy defects are
delineated by solid lines.} \centering{}\label{fig:Figure-Vacancy}
\end{figure}

The GGA band gap of a defect-free, 2D monolayer ZnO increases by
$\sim$0.08 eV in a (7$\times$7) supercell including a vacancy
defect. This is an artifact of the model, which mimics an
individual vacancy in supercell having limited size. First we
generated a Zn-vacancy by removing a single Zn atom from the
monolayer of ZnO in honeycomb structure as shown in
Fig.~\ref{fig:Figure-Vacancy} (a). Upon the structure optimization,
two O atoms around vacancy are departed from the plane. Similar to
the vacancies in graphene and BN, Zn-vacancy induces a local
magnetization in the system. Isovalue surfaces of spin density
difference $\Delta \rho$ clearly reveals the magnetism around
vacancy. The calculated total magnetic moment is 2 $\mu_{B}$ per
unit cell. Note that similar Zn-vacancy in the 3D bulk ZnO crystal
induces a magnetic moment of $\sim$ 1 $\mu_B$. It should be noted
that the magnetic moment of Zn-vacancy arises from the unpaired
electrons of oxygen atoms around the Zn vacancy. Therefore it is
important to determine precisely whether the vacancy induced
localized states are occupied. The positions of these localized
states of 2D ZnO calculated for the (7$\times$7) supercell and
their occupancy unambiguously comply with the calculated magnetic
moment. The Zn-vacancy in a repeating (7$\times$7) supercell also
modifies the electronic structure. The value of the band gap of
defect-free ZnO changes from 1.68 eV to 1.75 eV, and a spin up
localized state appear 0.25 eV above the top of the valence band.
Figure~\ref{fig:Figure-Vacancy} (b) presents our results for
O-vacancy. Unlike the case of Zn-vacancy, the monolayer of ZnO
containing an O-vacancy is nonmagnetic. Zn atoms around vacancy
with coordination number 2 prefer to stay in the same plane of the
other atoms and do not induce any magnetization. The band gap is
slightly modified to 1.80 eV. As for Zn+O divacancy in
Fig.~\ref{fig:Figure-Vacancy} (c), it is again nonmagnetic. The
band gap is also modified to 1.46 eV. Four occupied states
associated with divacancy occur as localized states in the band
gap. Finally, we consider the antisite defect. The resulting
relaxed structure is given in Fig.~\ref{fig:Figure-Vacancy} (d).
The antisite O is pushed away from Zn atoms and makes bonds with a
nearby O atom. The system does not show any magnetization. The
energy of final structure with antisite defect is $\simeq$ 5 eV
higher (i.e. energetically less favorable) that of the defect-free
2D monolayer of ZnO. The antisite induces two localized states in
the band indicated in the figure. Also it is noted that the
magnetic moments calculated for single Zn- and O-vacancy do not
agree with Lieb's theorem,\cite{lieb} which predicts the amount of
magnetic moments for carbon vacancies in 2D graphene. According to
Lieb's theorem, the net magnetic moment per cell is expected to be
$\mu$=1 $\mu_B$ for Zn and O vacancies in
Fig.~\ref{fig:Figure-Vacancy}. This might be related with the
ionic bonding which is different from graphene or the existence of
\textit{d-}orbitals in Zn. On the other hand, zero magnetization
for Zn+O divacancy in Fig.~\ref{fig:Figure-Vacancy} (c) is in
agreement with the theorem.

\subsection{2D ZnO bilayer}

Recently, Tusche \textit{et al.}\cite{bilayer} revealed two monolayer of
ZnO grown on Ag(111) substrate using surface x-ray diffraction and
scanning tunneling microscopy. They also showed that the
transition to the bulk wz-ZnO structure occurs in the 3-4
monolayer coverage. The Zn-O bond length of the planar hexagonal
structure measured 1.93 \AA{} is slightly larger than the the
value of 1.895 \AA{} calculated for the the bond length of the
monolayer of ZnO. This situation implies that the effect of the
Ag(111) substrate may be negligible.\cite{bilayer}

We investigated the atomic structure and stability of bilayer ZnO
honeycomb structure. To determine the minimum energy configuration
we used periodic supercell geometry and considered five different
stacking types which are $T_{1}$, $T_{2}$, $H_{1}$, $H_{2}$ and
$H_{3}$. In $T_{1}$ configuration, Zn (O) atoms of the second
layer are on top of the Zn (O) atoms of the first layer.  $T_{2}$
is similar to $T_{1}$ except that Zn atoms are above O atoms.
$H_{1}$, $H_{2}$ and $H_{3}$ configurations are obtained by
shifting one of the layers of $T_{1}$ and $T_{2}$ on the other
layer, so that Zn or O atoms of the second layer are placed above
the center of the hexagons in the first layer. It is $H_{1}$
($H_{2}$) if O (Zn) atoms of both layers face the centers of
hexagons.  $H_{3}$ corresponds to the configuration where O (Zn)
atom of the first (second) layer face the center of second (first)
layer.

Among these five configurations we found that $T_{2}$ is
energetically most favorable. $H_{3}$,  $H_{2}$, $H_{1}$ and
$T_{1}$ configurations have respectively 0.213, 0.312, 0.320 and
0.321 eV (per primitive cell) higher energies than $T_{2}$
configuration. The length of Zn-O bonds as well as the charge
transfer from Zn to O in $T_{2}$ configuration is slightly larger
than the value of 1.895 \AA{} calculated for the the bond length
of the monolayer of ZnO. This trend complies with the above
discussion that the bond length increases by going from 2D
monolayer to 3D bulk and implies that the effect of the Ag(111)
substrate may be negligible.\cite{bilayer}

The bilayer binding energy among two ZnO layers is calculated to
be 0.162 eV (per Zn-O pair) by subtracting the total energies of two individual
monolayers from the total energy of bilayer ZnO in $T_{2}$
configuration. Furthermore the layer-layer separations are
calculated as 4.02, 2.36, 3.80, 3.78 and 2.68 \AA{} for $T_{1}$,
$T_{2}$, $H_{1}$, $H_{2}$ and $H_{3}$ respectively. Hence, bilayer
formation is exothermic. The calculations with LDA, which
accommodate van der Waals interaction between layers better,
yields 0.297 eV (per Zn-O pair) binding energy between two ZnO layers and 2.267
\AA{} layer-layer separation in $T_{2}$ structure. The layer-layer
separation for $T_{2}$ is experimentally reported as 2.314
\AA{}.\cite{bilayer} The difference between the lattice constants
of the monolayer 2D ZnO honeycomb structure and the ZnO bilayer is
small. Owing to the relatively weak coupling between two ZnO
monolayers, the calculated electronic structure is similar to that
of single layer ZnO honeycomb structure, except that the band gap
decreases to 1.44 eV for $T_{2}$. This band gap increases to 5.10
eV after the \textit{GW$_{0}$} correction, which is still smaller than the
\textit{GW$_{0}$} corrected band gap of 2D monolayer ZnO.

We next address the question whether the structure of 2D bilayer
formed by the $T_{2}$ stacking of the ZnO bilayer is stable is
examined by the finite temperature ab-initio molecular dynamics
and phonon frequency calculations. Ab-initio molecular dynamics
calculations at 750 K are performed with the same parameters as
done for the monolayer ZnO in previous section indicate that the
bilayer remains stable at 750 K after 3.0 ps.

\begin{table*}
\begin{center}
\begin{tabular}{c|c|c|c|c|c|c|c|c}
ZnO & Bonding & $d_{Zn-O} \AA $ & $E_C$ (eV/atom) & GGA & $G_0W_0$
& \textit{$GW_0$} & \textit{$GW$} & Exp \\\hline
3D $wz$ & $sp^3$ & 2.001-2.007 & 8.934 & 0.75 & 2.76 & 3.29 & 2.44 \cite{hamada} & 3.37 \cite{bulk_zno}\\
3D $zb$ & $sp^3$ & 2.001 & 8.919 & 0.65 & 2.53 & 3.09 & 3.59 \cite{zb_gw} &  \\
3D $h$  & $sp^2+p_z+VdW$ & 1.999 & 8.802 & 0.96 & 2.84 & 3.32 &  &  \\
2D bilayer & $sp^2+p_z+VdW$ & 1.932 & 8.640 & 1.44 & 4.45 & 5.10 &  &  \\
2D monolayer & $sp^2+p_z$ & 1.895 & 8.478 & 1.68 & 4.87 & 5.64 & 3.57 \cite{phonon} & \\
\end{tabular}
\end{center}
\caption{A comparison of bonding (or type of hybrid orbitals), the
length of Zn-O bond $d_{Zn-O}$, cohesive energy per Zn-O pair
$E_C$, band gap $E_G$ calculated with GGA and corrected by
$G_{0}W_0$ and \textit{$GW_0$} in the present study, earlier \textit{$GW$}
calculations and experimental value.} \label{tab:bare}
\end{table*}

\begin{figure}
\begin{center}
\includegraphics[width=8cm]{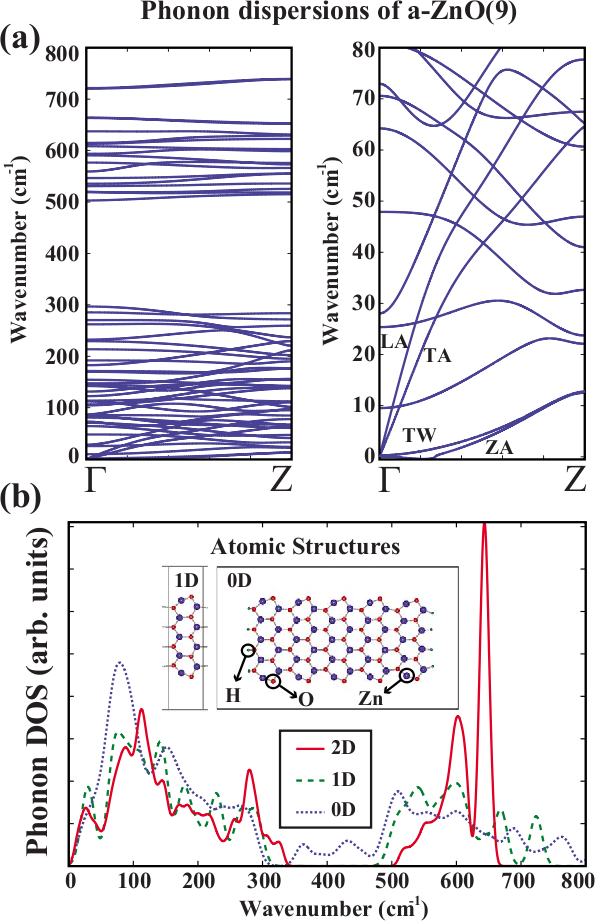}
\caption{(Color online)  (a) Phonon dispersion curves of
a-ZnO($9$)calculated by force constant method. Right panel
presents the magnified view of the low lying modes. Here the well
known out of plane, transverse and longitudinal acoustical modes
are labeled, respectively, as ZA, TA and LA. The fourth
acoustical twisting mode, which is indigenous to quasi 1D
nanoribbons, is labeled as TW, because this mode corresponds to
the twisting of the nanoribbon along the axis passing through the
middle of the ribbon in the infinite direction. (b) Density of
states (DOS) calculated for 2D planar ZnO (shown by solid lines);
armchair a-ZnO(9) (shown by dashed lines); a finite size flake,
i.e. a 5 unit cell long a-ZnO(9)(shown by dotted lines). The
discrete frequencies due to finite size structure are gaussian
broaden with $\sigma$=8 cm$^{-1}$ in order to compare with
continuous DOS of 2D layer and quasi 1D ribbon. Quasi 1D
(nanoribbon) and 0D (flake) honeycomb structures are shown by
inset.} \label{fig:Figure-1d-Phonon}
\end{center}
\end{figure}

\section{Dimensionality Effects}

A comparison of Zn-O bond length, cohesive energy, GGA bad gap,
\textit{GW$_{0}$} corrected band gap calculated for 3D wz-, zb-, h-ZnO, 2D
bilayer and monolayer of ZnO are presented in Table-I, where
interesting dimensionality effects are deduced. These
dimensionality effects are believed to be crucial for better
understanding of 2D crystals. Quasi 1D nanoribbons are not
included to this discussion because of edge effects of the ribbon.

Three dimensional crystals have larger number of nearest neighbors
and also posses larger Madelung energy as compared to 2D crystals.
wz-ZnO appears to correspond to the global minimum of ZnO II-VI
compound. However the energy difference between wz-ZnO and 2D
monolayer of ZnO is only $\sim$0.5 eV and is smaller than one
expects. It appears that $sp^2$-like bonding which is stronger
than $sp^3$-like bonding and the $\pi$-bonding between adjacent
$p_{z}$ orbitals, which contributes to stability by maintaining
the planar geometry give rise to the relatively smaller energy
difference between 3D and 2D monolayer ZnO. In the absence of
significant $\pi$-bonding one expects that 2D planar ZnO would be
buckled (puckered) for stability as found in 2D Si and GaAs
honeycomb structure.\cite{seymur,hasan} Through puckered planar
$sp^2$ orbital is dehybridized and is transformed to $sp^3$-like
hybrid orbitals. Here we note that population of oxygen $sp^2$ and
$p_z$ orbitals is larger than that of corresponding Zn orbitals,
since the former significantly higher electronegativity.

Interestingly, while GGA bad gaps of 3D ZnO occurs 0.75-0.96 eV,
the band gaps of in 2D is 1.68-1.44 eV. The band gap in 2D is
larger than 3D, since the energy difference between $sp^3$-like
orbitals of Zn and O ions is smaller than the energy difference of
$sp^2$-orbitals. The latter gives rise to larger energy difference
between bonding and antibonding orbitals. These band gaps are,
however, underestimated by GGA, since they increase approximately
three times upon \textit{GW$_0$} corrections. It appears that band gap
values corrected by \textit{GW$_0$} is closer to experimental value and
hence superior to \textit{GW} correction. As for as $d_{Zn-O}$, $E_C$ and
$E_G$ are concerned, 3D h-ZnO is intermediate between 3D and 2D
structures. Also calculated values of bilayer ZnO are slightly
closer to 3D than those of 2D monolayer of ZnO.

Charge transfer from Zn to O is crucial for dimensionality
effects, but difficult to calculate precisely. Charge transfer,
$\delta q$, calculated using three different schemes, namely
L\"{o}wdin, Bader and Siesta, yield different but consistent
values. For example, the L\"{o}wdin values occur around 1.4
electrons for 3D, but relatively smaller value of 1.35 electrons
for 2D. Bader analysis yields 1.22-1.20 electrons for 3D, and
relatively smaller value of 1.18 electrons for 2D. As for Siesta,
3D values, they are 0.90 electrons for 3D, but 0.87 electrons for
2D. Excluding the paradoxical situation with charge transfer
values of zb-ZnO occurring close to those of 2D ZnO, the charge
transfer between Zn and O is slightly smaller in 2D than 3D.

\section{ZnO Nanoribbons}

In this section, we consider bare and hydrogen passivated armchair
(a-ZnO) and zigzag (z-ZnO) ZnO nanoribbons. These nanoribbons are
specified according to their width given in terms of $n$ number of
Zn-O pairs in their unit cells. Hence, z-ZnO($n$) indicates a
zigzag ZnO nanoribbons having $n$ Zn-O pairs in their unit cell.
We investigate their electronic, magnetic and mechanical
properties. First, we start with the stability analysis of these
nanoribbons.

\subsection{Stability Analysis}

We have analyzed the stability of the bare armchair ZnO
nanoribbons having $n=9$ Zn-O atom pairs in the unit cell using
the force constant method and the finite temperature molecular
dynamics calculations. The dynamical matrix was generated using
the forces calculated in seven unit cells of a-ZnO(9). Results of
this calculation were presented in Fig.~\ref{fig:Figure-1d-Phonon}
(a). Phonon dispersion curves of quasi 1D nanoribbons, in general,
show profiles expected from the folding of 2D phonon dispersion
curves. Modes appearing above 700 $cm^{-1}$, however, are not
expected from this folding. They were attributed to edge effects
and reconstructions.

Right panel of Fig.~\ref{fig:Figure-1d-Phonon} (a) presents the
acoustic region in ten times magnified scale. There are four
acoustic modes, dispersion curves of which go to zero as
\textbf{k} $\rightarrow$ 0. To get exactly zero value, we have
imposed the symmetries originating from translational and
rotational invariance on the dynamical matrix. To impose these
symmetries we have tuned the force constant matrix elements so
that all forces on all atoms are zero when the whole structure is
shifted in three dimensions or slightly rotated along the axis
passing through the middle of the nanoribbon.

Both the longitudinal and transverse acoustic modes have linear
dispersions near the $\Gamma$ point. Group velocity of LA mode is
slightly higher than that of TA mode. Out of plane ZA and twisting
acoustical TW modes\cite{twisting} have quadratic dispersion near
the $\Gamma$ point, which is attributed to the rapid decay of the
force constants with increasing neighbor distance. In fact there
are some imaginary frequencies in ZA mode near the $\Gamma$ point.
The absolute value of these frequencies do not exceed 0.5
$cm^{-1}$ and they are purely an artifact of the precision of the
numerical calculation. Using a finer mesh makes these imaginary
frequencies disappear.

We also have calculated vibrational modes of the finite patch of
ZnO having the length of five unit cells and the width of $n=9$.
Zigzag edges of this flake were saturated by hydrogen atoms to
eliminate the magnetism and to simplify the numerical
computations. Vibrational spectrum of this structure had no
imaginary frequencies implying the fact that finite size flakes of
2D ZnO honeycomb structure is stable. The density of states, DOS of
calculated phonon frequencies are presented in
Fig.~\ref{fig:Figure-1d-Phonon} (b). Note that, DOS calculated for
three different honeycomb systems are similar. Owing to the edge
effects the gap between the acoustical and optical branches of the
ribbon is reduced. Similar effect as well as broadening of
discrete mode frequencies cause DOS of the flake to deviate
significantly from that of 2D honeycomb structure in the gap. The
vibrational modes attributed to strong Zn-H and O-H bonds are
centered respectively at 1830 $cm^{-1}$ and 3700 $cm^{-1}$. These
modes are not shown, since these frequencies are beyond the range
of Fig.~\ref{fig:Figure-1d-Phonon} (b).

Ab initio molecular dynamic calculations were also carried out for
a-ZnO(9) and a finite size patch of it. To eliminate constraints
to be imposed by small unit cell, the infinite nanoribbon is
treated by a supercell composed of five unit cells. Both
nanoribbons are kept at 750 K for 3.5 ps. As a result, infinite,
periodic structure of a-ZnO(9) composed of five unit cells had
minor reconstructions at the edges, while its honeycomb structure
was preserved. Finite structure had the similar pattern at the
armchair edges, but the reconstructions at the hydrogen saturated
zigzag edges were more pronounced. These reconstructions made the
whole system bend, but again the honeycomb structure was preserved
around the central region of the nanoribbon. The results of this
analysis are interpreted that finite size ZnO nanoribbons are
stable.

\begin{figure}
\begin{centering}
\includegraphics[width=7.8cm]{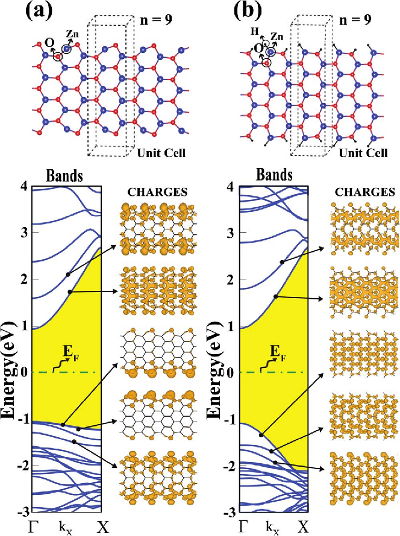}
\par\end{centering}
\caption{(Color online) Atomic and electronic structure of bare
and hydrogen terminated armchair nanoribbons a-ZnO($9$). (a)
Atomic structure with unit cell comprising $n$=9 Zn-O atom pairs,
energy band structure and charge density isosurfaces of selected
states of bare nanoribbons. The charge of the bands of edge states
are localized at the edge atoms. (b) Same as (a) but for hydrogen
passivated nanoribbon. Large medium and small balls indicate Zn, O
and H atoms, respectively. Unit cells are delineated by dashed
lines. } \centering{}\label{fig:Figure-Armchair}
\end{figure}

\subsection{Electronic and Magnetic Properties}

Bare and hydrogen terminated a-ZnO nanoribbons are nonmagnetic
semiconductors with direct band gaps which are relatively larger
than that of the monolayer ZnO. Fig.~\ref{fig:Figure-Armchair} (a)
and (b) shows the atomic and electronic structure of bare and
hydrogen terminated a-ZnO($n$) with $n$=9. The atoms at the edges
of the bare a-ZnO are reconstructed; while one edge atom, Zn is
lowering, adjacent edge atom, O is raised. The energy band gap
calculated with GGA is direct and 1.92 eV wide and is larger than
the band gap of 2D monolayer ZnO. Band decomposed charge density
analysis of a-ZnO(9) is presented in
Fig.~\ref{fig:Figure-Armchair}. The highest valance bands are
degenerate and their charge accumulates on oxygen edge atoms for
bare ribbon. The third band from the top of valence band is also
associated with edge states. On the other hand, the charge of the
lowest conduction band is distributed uniformly in the ribbon. The
charge of the second conduction band is mostly distributed at the
dangling bonds on the edges of the ribbon.

The passivation of Zn and O atoms at the edges by hydrogen atoms
gives rise to changes in the electronic band structure. The bond
lengths Zn-H and O-H bonds are calculated to be 1.53 and 0.97 \AA{}.
As seen in Fig.~\ref{fig:Figure-Armchair} (b), upon passivation
with hydrogen atoms, the reconstruction of edge atoms are
weakened. At the end the edge state bands are discarded from the
band gap and replaced by dispersive bands having charge
distributed uniformly in the ribbon. The band gap of H-passivated
a-ZnO(9) slightly increases to $\sim$1.98 eV.

The variation of the band gap $E_{G}$ for the bare and hydrogen
saturated armchair nanoribbons as a function of $n$ is given in
Fig. \ref{fig:Figure-BandGaps}. The band gaps are large for small
$n$ but approaches to those of 2D honeycomb structure as the width
$n \rightarrow \infty$. This is an indication of the quantum size
effect. For $n$<9, the value of the band gap of hydrogen
passivated a-ZnO is significantly larger than that of bare
ribbons; the difference practically disappears for $n$>20. In
contrast to graphene and BN nanoribbons, family dependent
variations of band gap\cite{family-dep-gaps} is absent in Fig.
\ref{fig:Figure-BandGaps}. The variation of the band gap with $n$
is an important property, which may lead to formation of quantum
dot or multiple quantum wells through the size
modulation.\cite{gribbon4}

Earlier it has been reported\cite{mendez1, mendez2} that all a-ZnO
nanoribbons are semiconductors with a constant band gap of 2 eV.
In our case, however, the band gaps of nanoribbons,  especially
for n<10,  display apparent dependency on n. The tendency of the
decrease in the band gaps to the 2D ZnO band gap is also observed
when n is increased.

\begin{figure}
\begin{centering}
\includegraphics[width=7.8cm]{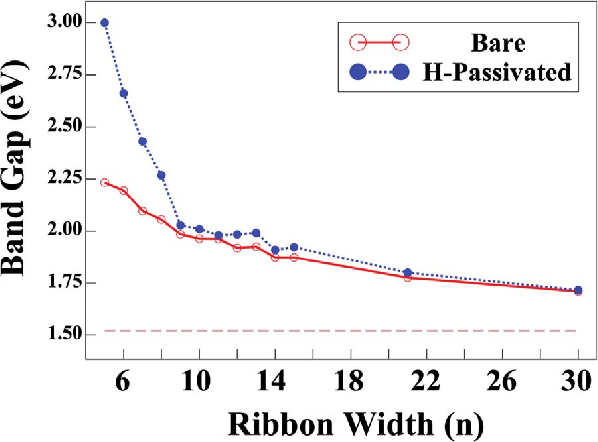}
\par\end{centering}
\caption{(Color online) Variation of the calculated band gaps of
bare and hydrogen passivated ZnO armchair nanoribbons with their
widths $n$. Dashed line indicates the band gap of the infinite 2D
ZnO.} \centering{}\label{fig:Figure-BandGaps}
\end{figure}

In contrast to a-ZnO nanoribbons, all zigzag nanoribbons (z-ZnO)
we investigated in this work (namely n=5,7,9) show metallic
character regardless of their widths. On the other hand, while all
a-ZnO nanoribbons are nonmagnetic, bare z-ZnO nanoribbons are
magnetic due to edge states. The magnetic properties of z-ZnO
depend on whether the edge atoms are passivated with hydrogen. Our
results regarding with the bare and hydrogen passivated z-ZnO($n$)
nanoribbons for $n$=9 are presented in
Fig.~\ref{fig:Figure-Zigzag} (a)-(c). In
Fig.~\ref{fig:Figure-Zigzag} (a), while the upper edge of the
ribbon is made by O atoms, the lower edge terminates with Zn
atoms. The bare z-ZnO nanoribbon is magnetic, since the spin
polarized calculations yield the total energy, which is
energetically 34 meV favorable than that obtained from spin
unpolarized calculations. The total magnetic moment of the system
was calculated as 0.57 $\mu_B$ per unit cell.
Figure~\ref{fig:Figure-Zigzag} (b) shows total density of states
(DOS) and band structure of bare z-ZnO(9) together with isosurface
charge densities of selected bands. Bare z-ZnO(9) have isosurfaces
of difference charge density $\Delta \rho$ occur around O edge
atoms due to unpaired \textit{O-$2p$ } orbitals. Clearly, bare
z-ZnO is a ferromagnetic metal. These results are in agreement
with those predicting that the ferromagnetic behavior of ZnO
nanoribbons due to unpaired spins at the edges is dominated by
oxygen atoms.\cite{mendez1, mendez2}

\begin{figure*}
\begin{centering}
\includegraphics[width=15cm]{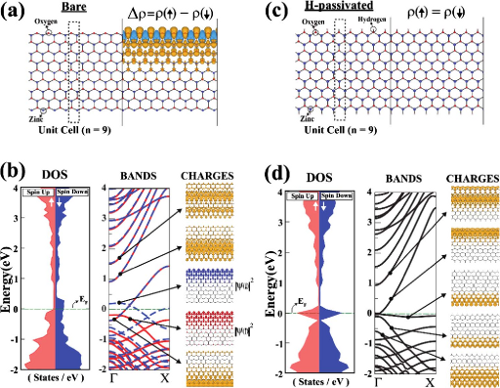}
\par\end{centering}
\caption{(Color online) Bare zigzag nanoribbon z-ZnO($n$) with
$n$=9: (a) Honeycomb structure with primitive unit cell delineated
and difference charge density $\Delta\rho$ of spin-up and
spin-down states.  (b) Total Density of spin-up and spin-down
states (DOS), energy band structure and isosurface charge density
of selected spin states. Large, medium and small filled balls
indicate Zn, O and H atoms, respectively. (c) and (d) are same as
(a) and (b). Yellow-light and blue-dark densities are for spin-up
and spin-down states. Similarly red-continues and blue-dashed
curves are spin-up and spin-down spin bands. Black lines are for
nonmagnetic (spin-paired) bands. }
\centering{}\label{fig:Figure-Zigzag}
\end{figure*}

The band structure of the nanoribbon gives us further information
about the magnetism of the system. When we plot the spin-up and
spin-down bands we observe the splitting of these bands around
Fermi level. The conduction and lower valance bands are degenerate
for spin-up and spin-down states, some of which are shown with
yellow isosurface charge densities in the figure. We also plotted
charge density of two states of spin-split nondegenerate bands.
Spin-down state is half-filled, while spin-up state is totally
filled. Furthermore, their charges are confined to the O-edge of
the ribbon. The spin polarization at the Fermi level is calculated
in terms of density of spin up and spin down states,
$D(\uparrow,\downarrow,E_{F})$, namely
$P(E_{F})=|D(\uparrow,E_{F})-D(\downarrow,E_{F})|/[D(\uparrow,E_{F})+D(\downarrow,E_{F})]$
and is around 80 \% for spin-down states although the spin-up
states are majority spins. This suggests z-ZnO($n$) with high spin
polarization at $E_F$ can operate as spin valve device.

As we discussed in Sec.~\ref{vacancy}, the magnetism of Zn-vacancy
in the monolayer ZnO is directly related with O atoms having
unpaired \textit{$2p$} orbitals. Similarly, bare z-ZnO nanoribbons
have also magnetic states on O terminated edge of the ribbon. The
splitting of spin-up and spin-down bands around Fermi level is
removed when the edges of bare z-ZnO(9) in
Fig.~\ref{fig:Figure-Zigzag} (a) is passivated by hydrogen. The
resulting structure is again metallic but nonmagnetic. The
electronic properties of the z-ZnO(9) ribbon passivated with
hydrogen is presented in Fig.~\ref{fig:Figure-Zigzag} (c). Two
bands just below the Fermi level is localized on the Zn edge of
the ribbon and lowest conduction band is localized on the O edge
of the ribbon. Similar results are  also obtained for n=5 and n=7.
The situation regarding the magnetism of bare z-ZnO is somewhat
different from that of the bare zigzag graphene nanoribbons, which
have ferromagnetic coupling along the edges, but antiferromagnetic
coupling between the edges.\cite{gribbon1} Moreover, unlike the
case in z-ZnO nanoribbons, the magnetism of zigzag graphene
nanoribbons are not destroyed upon termination of edges with
hydrogen atom.

\begin{figure}
\begin{centering}
\includegraphics[width=7.8cm]{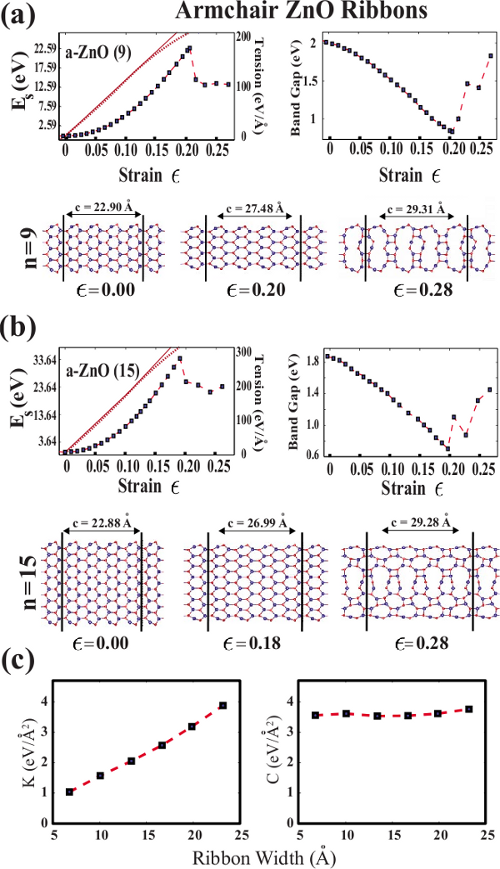}
\par\end{centering}
\caption{(Color online) (a) Variation of the strain energy $E_s$
and band gap $E_G$ of the bare a-ZnO($n$) ($n$=9) with the applied
uniaxial strain, $\epsilon$ are shown by a curve with squares.
Calculated tension  shown by tin (red) curve is initially linear,
but deviates from linearity for large $\epsilon$. Atomic structure
for three different values of strain. The boundaries of the
supercells comprising four unit cells and corresponding lattice
constants $c$ are indicated by vertical lines. (b) Same as (a) but
for bare a-ZnO($n$) ($n$=15). (c) Variation of force constant,
$\kappa$ and in-plane stiffness, $C$ of a-ZnO with the width $w$
of the nanoribbons.} \centering{}\label{fig:Figure-Kopmaa}
\end{figure}

An important feature of of zigzag ZnO nanoribbons is that charged
Zn and O atoms terminate different edges and thus induce sizable
electric dipole moments. For the bare z-ZnO(9) the dipole moment
is calculated to be 0.78 (electron$\times$\AA) per cell. Upon H
termination of Zn and O edge atoms the dipole moment increases to
1.07 (electron$\times$\AA) per cell and its direction is reversed. While
the dipole effects are included in the electronic structure
calculations of H-terminated z-ZnO(9) nanoribbons, the band gap
under estimated by GGA may effect its metallicity. Unfortunately,
\textit{GW$_0$} corrections cannot be applied due to large number of atoms.

\begin{figure}
\begin{centering}
\includegraphics[width=7.8cm]{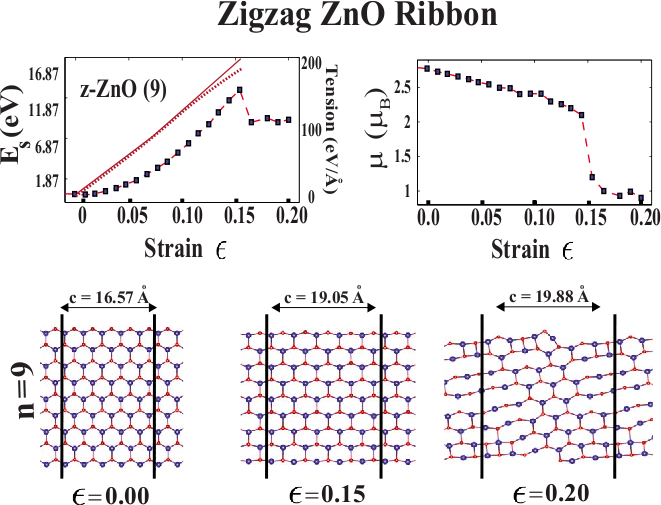}
\par\end{centering}
\caption{(Color online) Variation of the strain energy $E_s$,
tension and magnetic moment $\mu$ per supercell of the bare
z-ZnO($n$) ($n$=9) with the applied uniaxial strain. Atomic
structures of the bare z-ZnO(9) for three different uniaxial
strains. The supercells consist of five unit cells.}
\centering{}\label{fig:Figure-Kopmaz}
\end{figure}

\subsection{Mechanical Properties of ZnO Nanoribbons}

The response of the nanoribbons to the applied uniaxial stress is
crucial and provides information on the strength of the
nanoribbons. More recently, the response of graphene to strain (or
to tension) in the elastic deformation range has been an active
field of study. For example, recent works\cite{straineffect}
indicated the effect of deformation on the electronic properties
for band gap engineering. Moreover theoretical
studies\cite{mehmet3} have shown that carbon atomic chain can be
derived from graphene under tension. C. Jin \textit{et al.}\cite{carbonchain} 
showed that carbon atomic chains can be derived from graphene by electron irradiation 
inside a transmission electron microscope. Therefore, the response of 
ZnO nanoribbons to tension is of crucial importance.

Owing to ambiguities in defining the cross section of the ribbon
one cannot determine the Young's modulus rigorously. We examined
the variation of the strain energy,
$E_{s}(\epsilon)=E_{T}(\epsilon)-E_{T}(\epsilon=0)$ with respect
to the applied uniaxial strain, $\epsilon = \Delta c/ c$, $c$
being the lattice constant along the ribbon axis. The variation of
$E_{s}(\epsilon)$ includes information regarding the mechanical
properties of the ribbon. For example, force constant, $\kappa =
\partial^2E_T/\partial c^2$ is obtained from the variation of
$E_{s}$. $\kappa$ is an observable and can specify the strengths
of ribbons unambiguously. The effects of two edges due to
incomplete hexagons becomes important for narrow nanoribbons, but
decrease with increasing width. In-plane stiffness \textit{C},
which is independent of the thickness, can also be used instead of
Young's modulus. Defining $A_{0}$ as the equilibrium area of the
system, \textit{C} can be given as

\begin{equation}
\textit{C}=\frac{1}{A_{0}}(\frac{\partial^{2}E_{s}}{\partial\epsilon^{2}})
\end{equation}

By choosing a reasonable thickness ``\textit{h}'', Young's modulus
can be recalled as \textit{Y}=\textit{C}/h. The thickness value
around 3.34 \AA{} was used in order to evaluate the elastic
properties of SWNTs and  graphene by some works in
literature.\cite{yakobson1,yakobson2,dereli} We note that while
$C$ is unique for 2D infinite (periodic) honeycomb structure of
ZnO, for nanoribbons it depends on how the width of the ribbon is
taken in determining $A_{0}$. In fact, the width of the a-ZnO
nanoribbon cannot be determined straightforwardly. It is not clear
whether the distance is from Zn at one edge to Zn at the other
edge, or from O at one edge to other O. One can also take the
average of these two, or consider charge spill over from these
edge atoms. Hence the value of $C$ is subject to small changes
depending on how the width of the nanoribbon is taken. This
uncertainty, however, diminish as the width increases; eventually
the value of $C$ converges to a fixed value. In this respect, once
the width is fixed $C$, can be determined from $\kappa$.

We considered three nanoribbons, a-ZnO(9), a-ZnO(15) and z-ZnO(9).
In order to eliminate the constraints imposed by the periodic
boundary conditions of small unit cell, calculations are performed
using a supercell composed of four primitive unit cells for a-ZnO
and five unit cells for z-ZnO. The size of supercells is
determined based on certain tests. Figure~\ref{fig:Figure-Kopmaa}
(a) and (b) show the variation of the strain energy
$E_{s}(\epsilon)$ and band gap $E_{G}(\epsilon)$ with the applied
strain, $\epsilon$ for both ribbons, a-ZnO(9) and a-ZnO(15). They
display a parabolic $E_{s}$ vs $\epsilon$ curves up to the strains
$\epsilon \sim$ 0.13.  Beyond these strain values, energy vs
strain variation becomes elastic but anharmonic. In the elastic
range, the ribbons preserve their honeycomb like structure, but
the hexagons are elongated uniformly along the direction of the
strain. Zones of harmonic and anharmonic variation of
$E_{s}(\epsilon)$ can be better deduced by calculating the tension
force, $F_{T}(\epsilon)= -
\partial E_{s}(\epsilon)/\partial c$. $F_{T}(\epsilon)$ is linear
for $0 \leq \epsilon \leq \sim 0.13$ in the elastic-harmonic
range, but becomes nonlinear for $\sim 0.13 \leq \epsilon$ as seen
in Fig.~\ref{fig:Figure-Kopmaa}(a) and (b). If the applied tension
is released, the elastic deformation in both harmonic and
anharmonic ranges disappears and the ribbon returns to its
original equilibrium geometry.

The elastic deformation range of a-ZnO(9) and a-ZnO(15) ends with
a sharp fall of the total energy. This point is identified as the
yielding point of the ribbon occurring at $\epsilon\simeq$ 0.20.
The yielding points are followed with structural transformation,
where honeycomb cage structure starts to undergo a change and part
of the strain energy is relieved. Once the yielding point is
passed the ribbon can stretch under significantly low tension
and some fraction of the deformation will be permanent and
irreversible if the tension is released. Similar behavior occurs when
a nanowire of metals, such as Au or Cu; following an
order-disorder transformation the elastic deformation ends and the
wire is elongated by one lattice constant (or it deforms
plastically). However, in contrast to the present case, the
nanowire attains again the ordered state and start to deform
elastically.\cite{mehrez}

The force constant is calculated to be $\kappa$=2.05  and 3.88
eV/\AA{}$^{2}$ for a-ZnO(9) and a-ZnO(15), respectively. The
increase of $\kappa$ from 2.05 eV/\AA{}$^{2}$ to 3.88
eV/\AA{}$^{2}$ is due to the increase of the width from $n$=9
($\sim$ 12.81 \AA{}) to $n$=15 ($\sim$ 22.65 \AA{}). Note that if
the Hooke's law were valid for these nanoribbons, the ratio
$\kappa(15)/\kappa(9)$ would be equal to the ratio of widths,
namely 22.65/12.81=$\sim$1.77. The calculated value
(3.88/2.05=1.89) is slightly higher.  The discrepancy arises from
the edges of the ribbons, which respond to tension differently
from the central region. Therefore, deviation from the Hooke's law
becomes more serious as $n$ decreases, but diminish as $n
\rightarrow \infty$. $\kappa$ values of a-ZnO(9) and a-ZnO(15)
calculated for primitive unit cell are 7.92 eV/\AA{}$^{2}$
(instead of 4$\times$ 2.05=8.20 due to numerical calculations
performed in a large supercell) and 15.04 eV/\AA{}$^{2}$ (instead
of 4$\times$ 3.88=15.52), respectively. These values are smaller
than those corresponding to graphene ( 40.96 and 72.52
eV/\AA{}$^{2}$ ) and BN ( 34.02 and 60.46 eV/\AA{}$^{2}$ )
nanoribbons.

The calculated in-plane stiffness values $C$ for a-ZnO(9) and
a-ZnO(15) are 3.75, 3.71 eV/\AA{}$^{2}$,
respectively.\cite{stiffness} The calculated \textit{C} values of
armchair graphene (BN) nanoribbons for n=9 and n=15 are 18.45
(14.55) and 18.85 (14.66) eV/\AA{}$^{2}$, respectively. Graphene
and BN nanoribbons have significantly higher in-plane stiffness as
compared to a-ZnO nanoribbons. The difference mainly originate
from the fact that a-ZnO honeycomb structure has larger unit cell
than that of graphene and BN honeycomb structures. In view of the
calculated $\kappa$ and $C$ values, both graphene and BN
nanoribbons are stiffer than ZnO nanoribbons.

Figure~\ref{fig:Figure-Kopmaa} illustrates also the atomic
structure of a-ZnO(9) and a-ZnO(15) nanoribbons for three
different values of applied strain. After the yielding point, the
honeycomb structure is destroyed and the polygons being smaller and
larger than hexagons form. In some cases, a net magnetic moment
can develop in those polygons. The onset of plastic range and
variation of atomic structures can depend on the periodic boundary
conditions. These artifacts of periodic boundary conditions are
eliminated to some extent by using larger supercells comprising
several unit cells. Nonetheless, the plastic deformations shown in
Fig.~\ref{fig:Figure-Kopmaa} (a) and (b) have still large repeat
periodicity. Even if the size and forms of polygons appear to be
hysteric, their further investigation is of fundamental interest.
The order-disorder structural transformation, as well as
dislocation generation appear to be absent in this study. The
latter may require a treatment of deformation by taking into
account very large $n$ and $c$. Moreover, as predicted for
graphene nanoribbons, one expects that the onset of yielding in
the presence of vacancy defects and also at high temperature can
occur at small strain.\cite{mehmet3} Nevertheless, the plastic
deformation in the present work is carried out under ideal
conditions, hence its stochastic nature is not taken into account.
The energy gap of the nanoribbon undergoes a significant change
under uniaxial tension. It decreases as $\epsilon$ increases and
becomes very small at the yielding point. Beyond the yielding
point the band gap increases again.

In Fig.~\ref{fig:Figure-Kopmaa} (c) we plot the calculated values
of force constant, $\kappa$ relative to the width $w$ of
a-ZnO($n$) taken as the distance between outermost atoms of both
edges. As expected $\kappa$ is proportional to $w$ and $\kappa(w)$
can be considered as linear except small deviations from linearity
due to edge effects. In fact, $\partial \kappa(w) /\partial w$
increases slightly with $w$. Whether a-ZnO($n$) has even or odd
$n$ also causes to a small deviation from single linear variation.
As expected, the variation of the in-plane stiffness $C$ with $w$
is not significant. However, $C$ changes significantly between two
subsequent values of $n$.

The zigzag z-ZnO(9) nanoribbon displays a variation
$E_{s}(\epsilon)$ similar to those of a-ZnO. The calculated value
of $\kappa$ is 4.27 eV/\AA{}$^{2}$, which is larger than that
calculated for a-ZnO(9) having relatively smaller width.
Similarly, in-plane stiffness value of z-ZnO(9) is slightly
smaller than that of a-ZnO(9) and a-ZnO(15). \textit{C} is
calculated as 3.24 eV/\AA{}$^{2}$. The zigzag ZnO(9) remains to be
ferromagnetic metal in the elastic deformation range, but their
magnetic moment decreases with increasing $\epsilon$. However,
beyond the yielding point it shows a sharp fall and diminish with
increasing $\epsilon$ in the plastic range as seen in
Fig.~\ref{fig:Figure-Kopmaz}. In contrast to a-ZnO, long polygons
are aligned along the axis of the z-ZnO(9) in
Fig.~\ref{fig:Figure-Kopmaz}.

\section{Discussion and Conclusions}

This work presents an extensive study on 2D monolayer and bilayer
ZnO, and ZnO nanoribbons together with their stability analysis.
The monolayer of ZnO is an ionic and nonmagnetic, wide band gap
semiconductor with significant charge transfer from zinc atoms to
nearest oxygen atoms. Since DFT normally underestimates the
calculated band gaps, the calculated band gap of specific
structures are corrected by \textit{GW$_{0}$} self-energy calculations. ZnO
has 2D hexagonal lattice forming a planar honeycomb structure. Our
predictions, which contribute to a better understanding of this
material are emphasized by way of conclusions:

i) We have shown that, 2D ZnO monolayer and bilayer, quasi 1D bare
nanoribbons of ZnO and 0D patch of ZnO correspond to local minima
on the Born-Openheimer surface and thus are predicted to be
stable. ii) Ab-initio molecular dynamics calculations performed at
high temperature corroborate our analysis obtained from the
calculation of phonon frequencies. iii) We performed calculations
of phonon modes and density of states of phonon frequencies of 2D
monolayer, quasi 1D armchair nanoribbon and a finite flake using
force constant method. In particular, we revealed the acoustical
twisting mode of armchair nanoribbon. iv) Our study of Zn-, O-,
Zn-O-vacancies and O-Zn antisite indicates that local magnetic
moments can be generated only by Zn-vacancies. v) We provided an
extensive analysis of the electronic structure of hydrogen
terminated, as well as bare armchair and zigzag nanoribbons.
Armchair ZnO nanoribbons are found to be nonmagnetic
semiconductors. The band gaps vary with their widths. The narrow
nanoribbons have relatively larger band gap due to the quantum
confinement effect. Bare zigzag ZnO nanoribbons are ferromagnetic
metals. The atoms near the oxygen terminated edge of the ribbons
acquire magnetic moments. The spin-polarization at the Fermi level
may attain high values for specific zigzag nanoribbons. However,
once O- and Zn- terminated edges are passivated with hydrogen
atoms, the zigzag nanoribbon becomes nonmagnetic metal. These
electronic and magnetic properties might be useful for spintronic
applications. vi) We found the minimum energy stacking of the ZnO
bilayer, and provided energetics and energy band structure
corresponding to this stacking. Accordingly, bilayer ZnO is also a
nonmagnetic, wide band gap semiconductor with slightly smaller
band gap as compared to that of monolayer. vii) We provided an
analysis of mechanical properties. ZnO nanoribbons under uniaxial
strain show harmonic and anharmonic elastic deformation ranges and
a yielding point. After yielding, the strain energy exhibits a
sharp fall and the nanoribbon deform plastically. In the elastic
range, hexagons are uniformly deformed, but honeycomb like atomic
structure is maintained. After yielding point some of the hexagons
are modified and reconstruct to different polygons which may be
smaller or larger than hexagons. Variation of electronic and
magnetic properties with  deformation and formation of polygons in
the plastic deformation range are of fundamental interest.
Calculation of force constants and in-plane stiffness indicate
that the stiffness of ZnO nanoribbons is smaller than those of
graphene and BN honeycomb structures.

In summary, it is shown that single and bilayer ZnO and its
nanoribbons in different orientations provide us for a variety of
electronic and magnetic properties which may be interesting for
further applications. Even if they have honeycomb structure common
to monolayer graphene and BN, their properties exhibit important
differences.

\begin{acknowledgments}
We acknowledge stimulating discussions with Dr. Ethem Akturk. Part
of the computations have been provided by UYBHM at Istanbul
Technical University through a Grant No. 2-024-2007. This work is
partially supported by TUBA, Academy of Science of Turkey.
\end{acknowledgments}


\begin{thebibliography}{15}
\bibitem{radius}
Calculated effective charge of Zn atom indicating $\sim$1 electron
occupancy of the valance orbitals eliminates the inert gas
configuration. Therefore, one considers the covalent radii, namely
R$_{Zn}$=1.31 \AA~and R$_{O}$=0.66 \AA.
\bibitem{family-dep-gaps}
The variation of $E_G$ with $n$ shows an family dependent behavior
for armchair nanoribbons of graphene, BN and even Si \cite{hasan}.
For example, the band gap is small for $n=3m-1$ ($m$ being an
integer), but from $n=3m$ to $n=3m+1$ it increases and passing
through a maximum it becomes again small at the next minimum
corresponding to $n=3m-1$ for armchair graphene nanoribbons.
\bibitem{stiffness} Because of uncertainties in measuring the width of a bare armchair
nanoribbon, the value of $C$ cannot be determined unambigiously. For example, by taking $w$
as the distance between Zn atoms at both edges, $w=$12.186 \AA; alternatively $w$=13.393 \AA~
if the width is taken as the distance between O atoms at both edges of a-ZnO(9). The
corresponding in-plane stiffness values are 3.977 eV/\AA{}$^{2}$ and 3.618 eV/\AA{}$^{2}$,
respectively. One obtain $C\cong$ 3.75 eV/\AA{}$^{2}$. In conclusions, while the value of $C$
can be unambigiously determined for 2D periodic ZnO honeycomb structure, the value of $C$ calculated
for a nanoribbon depends on its width. As $w$ increases, $C$ decreases, and eventually
as $w \rightarrow \infty$ $C$ approaches the value calculated for 2D periodic ZnO honeycomb structure.
\end{thebibliography}
\end{document}